\documentclass[11pt,twoside]{article}
\usepackage{atmp}
\usepackage{amsmath,amsgen,amstext,amsbsy,amsopn,amsthm,amssymb}

\copyrightnotice{2002}{6}{847}{871}

\newtheorem{thm}{THEOREM}

\newtheorem{lem}{LEMMA}

\theoremstyle{definition}

\newcommand{\R}{{\mathbb R}}
\newcommand{\C}{{\mathbb C}}

\newcommand{\eps}{\varepsilon}
\newcommand{\A}{{\mathcal A}}
\newcommand{\Q}{{\mathcal Q}}
\newcommand{\Ss}{{\mathcal S}}
\newcommand{\T}{{\mathcal T}}
\newcommand{\E}{{\mathcal E}}
\renewcommand{\H}{{\mathcal H}}
\newcommand{\F}{{\mathcal F}}
\newcommand{\Nb}{{\mathcal N}}

\newcommand{\B}{{\mathcal B}}

\newcommand{\x}{{x}}

\newcommand{\half}{\mbox{$\frac{1}{2}$}}

\newcommand{\mph}{m}
\newcommand{\as}{\sqrt{\alpha}}
\newcommand{\vs}{ \sigma}
\newcommand{\const}{{\rm const. \,}}
\newcommand{\ean}{\varepsilon_{\lambda}}

\numberwithin{equation}{section}

\begin{document}

\title{Mass~Renormalization~and~Energy Level~Shift~in~Non-Relativistic~QED}

\url{math-ph/0205044}

\author{Christian Hainzl$^1$, Robert Seiringer$^2$}
\address{$^1$Mathematisches Institut, LMU M\"unchen\\ 
Theresienstrasse 39, 80333 Munich, Germany \\ 
hainzl@mathematik.uni-muenchen.de \\ \vskip .3cm
$^2$Department of Physics, Jadwin Hall, Princeton University\\
P.O. Box 708, Princeton, New Jersey 08544, USA\\ 
rseiring@math.princeton.edu}

\markboth{\it MASS RENORMALIZATION AND ENERGY LEVEL SHIFT\ldots}{\it C. HAINZL, R. SEIRINGER}

\begin{center}
{\it Dedicated to Elliott Lieb on the occasion of his 70th birthday}
\end{center}

\begin{abstract}
Using the Pauli-Fierz model of non-relativistic quantum
electrodynamics, we calculate the binding energy of an electron in the
field of a nucleus of charge $Z$ and in presence of the quantized
radiation field. We consider the case of small coupling constant
$\alpha$, but fixed $Z\alpha$ and ultraviolet cut-off $\Lambda$. We
prove that after renormalizing the mass the binding energy has, to
leading order in $\alpha$, a finite limit as $\Lambda$ goes to
infinity; i.e., the cut-off can be removed. The expression for the
ground state energy shift thus obtained agrees with Bethe's formula
for small values of $Z\alpha$, but shows a different behavior for
bigger values.
\end{abstract}

\footnotetext[1]{Marie Curie Fellow} 
\footnotetext[2]{Erwin Schr\"odinger Fellow. On leave from 
Institut f\"ur Theoretische Physik,
Universit\"at Wien, Boltzmanngasse 5, A-1090 Vienna,
Austria}

\cutpage
\section{Introduction}

Quantum electrodynamics (QED) has proved one of the most successful
theories in physics. One of its striking features was the explanation
of the so-called Lamb-shift, the splitting of the energy levels of the
hydrogen atom. Most of the results, however, are of perturbative
nature, and very little is known concerning rigorous results starting
from a well-defined (Hamiltonian) theory.

A few years ago, Bach, Fr\"ohlich and Sigal initiated a rigorous
non-pertur\-bative study of non-relativistic QED. They made a detailed
spectral analysis of the Pauli-Fierz Hamiltonian, which is a model
appropriate for studying the low energy properties of QED, and which
is the model we use in this paper. Among other results, they proved in
\cite{BFS} the existence of a ground state for small values of the
coupling constant $\alpha$ and the ultraviolet (UV) cut-off
$\Lambda$, which has to be introduced in the interaction term between the
electrons and the photon field in order to obtain a well-defined
Hamiltonian.  Recently, Griesemer, Lieb and Loss
\cite{GLL} succeeded 
in removing these restrictions on the parameters.  We refer to
\cite{GLL} for an extensive list of references on the subject.

Our goal in this paper is to calculate the binding energy of one
electron in the field of a nucleus of charge $Z$. In contrast to the
case of a Schr\"odinger operator without coupling to the quantized
photon field, the \lq\lq bare mass\rq\rq\ appearing in the Hamiltonian
is not equal to the physical mass. To obtain the binding energy as a
function of the physical mass, which can be measured in experiments,
and hence to compare the calculated binding energy with the measured
one, one has to substitute the physical mass for the bare mass. To do
this, one first has to calculate the physical mass as a function of
the bare mass, the coupling constant $\alpha$, and the ultraviolet
cut-off $\Lambda$. We do this via the dispersion relation,
i.e., the energy of a free electron as a function of the total
momentum (= electron momentum + field momentum).

There are three parameters in the problem: the coupling constant
$\alpha$, the UV cut-off $\Lambda$, and the nuclear charge $Z$. We
consider the case when $\alpha$ is small, but $Z\alpha$ not
necessarily small. I.e., we first fix $\Lambda$ and
$Z\alpha$ and consider $\alpha$ as a small parameter. 

Our calculations are to leading order in $\alpha$. We give rigorous
error estimates, but do not focus on optimizing the error terms. In
particular, we do not have sufficient control on their behavior for
large $\Lambda$ to be able to prove the finiteness of the result after
removing the cut-off beyond leading order in $\alpha$. This remains an
interesting open problem. Steps in this direction were taken in
\cite{LL}, where the behavior of the self energy for large $\Lambda$
was studied. Bounds on mass renormalization and binding energies were
recently given in \cite{LL2}.

We shall rigorously show that after renormalizing the mass the UV
cut-off can be removed, at least to leading order in $\alpha$.  More
precisely, the binding energy, as function of the physical mass and
the coupling parameter $\alpha$, has, to leading order in $\alpha$, a
finite limit as $\Lambda\to\infty$. We thus obtain an expression for
the ground state energy shift due to the presence of the radiation
field, which holds to leading order in $\alpha$, and {\it for all
values} of $Z\alpha$. It agrees with Bethe's formula
\cite{Be} for small values of $Z\alpha$, but shows a
different behavior for bigger values. We emphasize again that in our formula
for the binding energy no UV cut-off appears, in contrast to Bethe's
formula, who used the dipole approximation for it's derivation and
neglects the spin of the electron. In Bethe's formula a logarithmic
dependence on $\Lambda$ appears. Bethe argues that a
physically reasonable value for $\Lambda$ is $mc^2$, the energy needed
for pair production of electrons and positrons to take place. By
inserting this value for $\Lambda$ in his formula he obtains excellent
agreement with experiments, at least for the case $Z=1$, i.e., the
hydrogen atom.

Shortly after Bethe's work, it was noted in \cite{kroll} that the Lamb
shift ought to be finite after removing the cut-off $\Lambda$ if one
does not use the dipole approximation (see also \cite{au}).  By means
of perturbation theory the Lamb shift was calculated, for both
$\alpha$ and $Z\alpha$ small, in
\cite{grotch}.

Our proofs below essentially show that second order perturbation
theory is correct. I.e., to obtain results up to order
$\alpha$ only the addition of {\it one} photon is needed. We use and
extend previous ideas in \cite{H1}, where results on the leading
order of the self energy and the binding energy were obtained. (Concerning the question of enhanced binding, see \cite{HVV}.)

The paper is organized as follows. In the next section, we describe
the precise setting and introduce some notation that will be used
throughout the paper. In Section~\ref{massrensect} we calculate the
dispersion relation and thus obtain an expression for the physical
mass in terms of the bare mass, i.e., for the necessary mass
renormalization. The binding energy is calculated in
Section~\ref{bindsect}. In Section~\ref{rensect} it is shown that
after renormalization of the mass, the UV cut-off can be removed, and
a finite expression for the binding energy to leading order in
$\alpha$ is obtained. See Theorem~\ref{renbindthm} on
page~\pageref{renbindthm} below for the precise statement. Finally we
comment on the Lamb shift of metastable excited states in
Section~\ref{lambsect}.

\section{Setting and Notation}

We now describe the precise setting and introduce useful
notations. The Hamiltonian under consideration is the {\it Pauli-Fierz Hamiltonian}
\begin{equation}\label{rpf}
H = \frac 1{2m_0}\Big[\big(p + \sqrt{\alpha}A(\x)\big)^2 +
\as\vs\cdot B(\x)\Big] + V(\x) +  H_f\ .
\end{equation}
It acts on $\H=L^2(\R^3,d^3x;\C^2)\otimes \F_b(L^2(\R^3,d^3k;\C^2))$,
where $\F_b$ denotes the bosonic Fock space. Self-adjointness of $H$ on an
appropriate domain has recently been shown in \cite{hiro}.

Units are chosen such that $\hbar=c=1$. The electron charge is then
given by $e=\as$, with $\alpha$ the fine structure constant. In nature
$\alpha\approx 1/137$, but we consider it here as an arbitrary, small
number. The positive parameter $m_0$ is the bare mass of the electron.

The $\,\cdot\,$ product stands for the usual scalar product in $\R^3$,
and $\vs$ denotes the vector of Pauli matrices, acting on the spin
variable of the electron part of the wave function. The operator
$p=-i\nabla_\x$ is the electron momentum. The vector potential in
Coulomb gauge is
\begin{equation}
A(x) = \frac 1{2\pi} \sum_{\lambda = 1,2} \int_{|k|\leq \Lambda}
\frac{1}{\sqrt{|k|}} \ean(k) \Big[a_\lambda(k) e^{ik\cdot x} + a^*_\lambda (k) e^{-ik\cdot x}\Big] d^3k \ .
\end{equation}
Here the integration is restricted to momenta $|k|\leq \Lambda$, i.e.,
we make a sharp UV cut-off. We could allow for a more general,
spherically symmetric cut-off, but we restrict ourselves to this case
for simplicity. The vectors $\ean(k) \in \R^3$ are any two possible
orthonormal polarization vectors perpendicular to $k$. 
The corresponding magnetic field $B={\rm curl\, } A$ reads
\begin{multline}
B(x) = \frac 1{2\pi}\sum_{\lambda = 1,2} \int_{|k|\leq \Lambda}
\frac{i}{\sqrt{|k|}} \, k\wedge \ean(k)\,\Big[ a_\lambda(k) e^{ik\cdot x}- a^*_\lambda (k) e^{-ik\cdot x}\Big] d^3k \ .
\end{multline}
The operators $a^*_\lambda(k)$ and $a_\lambda(k)$ are the usual
creation and annihilation operators for a photon of momentum $k$ and
polarization $\lambda$. They satisfy the commutation relations
\begin{equation}
[a_\nu (k),a^*_\lambda(q)] = \delta(k-q)\delta_{\nu\lambda}
\end{equation}
and
\begin{equation}
[a_\nu^* (k),a^*_\lambda(q)] = 0\ , \quad
[a_\nu(k), a_\lambda(q)] = 0 \ .
\end{equation}
The field energy is given by
$H_f=d\Gamma(|k|)$, where we denote, in general, the second quantized version of a function $h(k)$ by
\begin{equation}
d\Gamma(h(k))= \sum_{\lambda = 1,2} \int_{\R^3} h(k) a_\lambda^*(k) a_\lambda(k)  d^3k \ .
\end{equation}

A general wave function $\Psi\in\H$ can be written as a sequence
\begin{equation}
\Psi=\Big\{ \psi_0, \psi_1,\psi_2,\dots \Big\} \ ,
\end{equation}
with $\psi_n=\psi_n(x,k_1,\dots,k_n)$ an $n$-photon wave function. We
suppress the dependence on the spin of the electron and the
polarization of the photons for simplicity of notation. The creation parts of the vector fields $A$ and $B$ create a photon with wave function $G(k)e^{-i k\cdot x}$ and $H(k)e^{-ik\cdot x}$, respectively. 
Here $G(k)=(G^1(k),G^2(k))$ and $H(k)=(H^1(k),H^2(k))$ are vectors of one-photon states, given by
\begin{equation}\label{defG}
G^{\lambda}(k)=\frac {\theta(\Lambda-|k|)}{2\pi |k|^{1/2}} \eps_\lambda(k) 
\end{equation}
and
\begin{equation}\label{defH}
H^{\lambda}(k)=\frac {-i\theta(\Lambda-|k|)}{2\pi |k|^{1/2}} k\wedge \eps_\lambda(k)  \ .
\end{equation}
Their vector components will be denoted by $H_i$ and $G_i$, respectively
($i=1,2,3$), and likewise generally for vectors in $\R^3$. The
function $\theta$ denotes the Heaviside step function, taking the
ultraviolet cutoff $\Lambda$ into account.

\section{Dispersion Relation}\label{massrensect}

We first consider the case of a free electron, i.e., $V=0$. The
Hamiltonian is then translation invariant, i.e., it commutes with the
three components of the total momentum $P=p+d\Gamma(k)$. It is
therefore possible to write the Hilbert space and the Hamiltonian as a
direct integral
\begin{equation}
\H=\int_{\R^3}^\oplus d^3P\, \H_P
\end{equation}
and
\begin{equation}
H=\int_{\R^3}^\oplus d^3P\, H_P \ ,
\end{equation}
with $H_P$ acting on $\H_P$. 
Each $\H_P$ is isomorphic to $\C^2\otimes\F_b$. The Hamiltonian $H_P$ in this representation is given by  
\begin{equation}
H_P = \frac 1{2m_0}\Big[\big(P-d\Gamma(k)+ \sqrt{\alpha}A(0)\big)^2 +
\as\vs\cdot B(0)\Big] +  H_f\ .
\end{equation}
For any fixed total momentum $P$, we define $\E_P$ to be the 
ground state energy of $H_P$. The existence of a corresponding ground state is of no concern to us. In \cite{chen} it was shown, at least for the spinless case, that there does not exist a ground state of $H_P$ for $P\neq 0$ unless one imposes an infrared cut-off on $H_P$.

The function $\E_P$ is often referred to as the {\it dispersion
relation}. For small $P$, $\E_P\approx\E_0+ |P|^2/2m$, where $m$ is,
by definition, the {\it physical mass}. It is strictly bigger than the
bare mass $m_0$. For given $\Lambda$, $m_0$ has to be chosen such that
$m$ is equal to the mass of the electron, which is determined by
experiment. The bare mass goes to zero as the UV cut-off $\Lambda$
goes to infinity, which is usually referred to as {\it
renormalization} of $m_0$.

We now calculate this mass renormalization to leading order in
$\alpha$. We restrict ourselves to the case $P<m_0$, since we are only
interested in the behavior of $\E_P$ for small $P$. For large $P$ the
behavior is different due to the different energy-momentum relations
of the relativistic photon and the non-relativistic electron.
 
\begin{thm}[{\bf Dispersion Relation}]\label{T1}
For $|P|< m_0$, the energy $\E_P$ of a free particle with total
momentum $P$ satisfies
\begin{multline}\label{ep}
\E_P-\E_0= \frac{|P|^2}{2m_0} - \frac{\alpha}{2\pi^2 m_0 } \int_{|k|\leq \frac \Lambda{2m_0}}
d^3k \frac{|k|}{|k|^2+|k|-P\cdot k/m_0}\times \\ \times \left(\frac {|P|^2}{|k|^2} + |P\cdot k|^2 \left(\frac 2{(|k|^2+|k|)^2} - \frac 1{|k|^4}\right)\right) 
+O(\alpha^{3/2}) \ .
\end{multline}
\end{thm}

Strictly speaking, we do not claim that the error term is truly
$O(\alpha^{3/2})$, but only that it is bounded by $\const
\alpha^{3/2}$ for small $\alpha$. 
The same remark applies to Theorems~\ref{uppevthm} and~\ref{bindthm}
below.

\begin{proof} Throughout the proof we will set $m_0=\half$ for simplicity. The correct dependence on $m_0$ is easily obtained by dimensional analysis. We will provide appropriate upper and lower bounds on $\E_P$. 

We start with the {\it upper bound:} We choose as a trial state
\begin{equation}\label{3.5}
\Psi=\left\{\uparrow, \frac {-\as}{|k|^2+|k|-2P\cdot k+\alpha}\Big[ 2 P\cdot G(k)+ \vs\cdot H(k) \Big]\uparrow ,0,0,\dots\right\} \ ,
\end{equation}
where $G(k)$ and $H(k)$ are defined in (\ref{defG}) and (\ref{defH}),
and where $\uparrow$ is a shorthand notation for the vector $(1,0)$ in
spin space. The additional factor $\alpha$ in the denominator of the one-photon part serves as an infrared cut-off to make the norm of $\Psi$ finite. 

 We write $A=A(0)=D+D^*$, where $D$ denotes the part of $A$
containing only annihilation operators.  A straightforward
calculation, using that
\begin{equation}
A^2=(D+D^*)^2= \frac 1\pi \Lambda^2 + 2 D^*\cdot D +D^*\cdot D^* + D\cdot D
\end{equation}
and $k\cdot D^* \uparrow=k\cdot G(k)\uparrow=0$, yields
\begin{multline}
\langle \Psi|H_P|\Psi\rangle = \left(|P|^2 + \frac \alpha\pi \Lambda^2\right) \langle \Psi|\Psi\rangle \\ - \alpha\Big\langle (2P\cdot G+\vs \cdot H)\uparrow \Big| \frac 1{|k|^2+|k|-2P\cdot k+\alpha} \Big|  (2P\cdot G+\vs \cdot H)\uparrow \Big\rangle \\ + \alpha^2 \big(\langle\Psi|\Psi\rangle-1\big)+ \const \alpha^2 \ .
\end{multline}
Using the fact the $G$ is real and $H$ is purely imaginary, and the anti-commutation relations for $\vs$, namely
\begin{equation}
\sigma_i\sigma_j+\sigma_j\sigma_i = 2 \delta_{ij} I_{\C^2} \ ,
\end{equation}
we get
\begin{multline}\label{38}
\Big\langle (2P\cdot G+\vs \cdot H)\uparrow \Big| \frac 1{|k|^2+|k|-2P\cdot k+\alpha} \Big|  (2P\cdot G+\vs \cdot H)\uparrow \Big\rangle = \\ \sum_{\lambda=1}^2 \int d^3k \frac { 4 |P\cdot G^\lambda(k)|^2 + |H^\lambda(k)|^2 } {|k|^2+|k|-2P\cdot k+\alpha} \ . 
\end{multline}
We now insert the expressions (\ref{defG}) and (\ref{defH}) for $G$ and $H$ and use that $\sum_{\lambda=1}^2 |P\cdot \eps_\lambda(k)|^2=|P|^2 - |P\cdot k|^2/|k|^2$ by the orthogonality relations of $k$ and $\eps_\lambda$. 
Moreover, since $\langle\Psi|\Psi\rangle = 1 + O( \alpha\ln(1/\alpha))$, we arrive at the upper bound
\begin{multline}\label{upbo}
\E_P\leq |P|^2 + \frac \alpha\pi \Lambda^2 - \frac\alpha{(2\pi)^2}  \int_{|k|\leq \Lambda}  d^3k \frac { 4 \left(|P|^2|k|^2- |P\cdot k|^2\right) + 2|k|^4 } {|k|^3(|k|^2+|k|-2P\cdot k)} \\ + O\big(\alpha^2\ln(1/\alpha)\big) \ .
\end{multline}

\noindent {\it Lower Bound:} 
We first show that, for suitable constants $C_i>0$,
\begin{equation}\label{apri}
H_P\geq \big(|P|^2+|d\Gamma(k)|^2\big)(1-C_1\alpha)+ C_2 H_f - C_3\alpha \ .
\end{equation}
Together with the upper bound (\ref{upbo}) this shows that, in an approximate ground state $\Psi_0$, $\langle \Psi_0|H_f|\Psi_0\rangle\leq \const \alpha$ and  $\langle \Psi_0|\,|d\Gamma(k)|^2\, |\Psi_0\rangle\leq \const \alpha$.

The fact that $\eps_\lambda(k)\cdot k=0$ implies that $d\Gamma(k)\cdot A=A\cdot d\Gamma(k)$. Using this we can estimate by Schwarz' inequality 
\begin{multline}\label{311}
\pm\as\langle \psi | (P-d\Gamma(k))\cdot A |\psi\rangle =\pm 2 \as\, \Re \, \langle \psi | (P-d\Gamma(k))\cdot D |\psi\rangle \\  \leq \alpha a \langle \psi | |P-d\Gamma(k)|^2 |\psi\rangle + \frac 1 a \langle \psi | D^*D |\psi\rangle
\end{multline}
for any positive constant $a$. This leads to the operator inequality  
\begin{equation}
\pm \as (P-d\Gamma(k))\cdot A\leq a\alpha |P-d\Gamma(k)|^2+\frac 1a D^* D \ .
\end{equation}
In the same manner, we obtain 
\begin{equation}
\pm\as \vs\cdot B\leq \frac1{a'} E^*E + \alpha a' 
\end{equation}
for a positive $a'$. Moreover, $D^*D\leq \mbox{$\frac 2\pi$} \Lambda
H_f$ and $E^*E\leq \mbox{$\frac 2{3\pi}$} \Lambda^3 H_f$ (see, e.g.,
\cite[Lemma~A.4]{GLL}; note the different prefactors due to the
missing $\mbox{$\frac 1{2\pi}$}$ in their definition of $A$). Since
$A^2\geq 0$ we get
\begin{multline}
H_P\geq |P-d\Gamma(k)|^2(1-2a\alpha)+H_f\left(1-\frac 4{a\pi} \Lambda - \frac 2{3a'\pi} \Lambda^3\right)-\alpha a' \\ \geq \big(|P|^2+|d\Gamma(k)|^2\big)(1-2a\alpha)+H_f\left(1-\frac 4{a\pi} \Lambda -  \frac 2{3a'\pi} \Lambda^3-2|P|\right)-\alpha a' \ ,
\end{multline}
where the last inequality holds for $\alpha\leq 1/(2a)$. Since $|P|<\half$ by assumption, we can choose the constants $a$ and $a'$ appropriately and therefore arrive at (\ref{apri}).

Using the results obtained above, we can now estimate $\langle\Psi_0 |H_P|\Psi_0\rangle$, for an approximate ground state $\Psi_0$, from below. By Schwarz' inequality
\begin{equation}
D^2+D^{*2}\geq - a D^*D - \frac 1 a DD^*
\end{equation}
for any $a>0$. Choosing $a=1/\sqrt{\alpha}>\half$ we get, using $DD^*=D^*D+\mbox{$\frac 1\pi$}\Lambda^2$, 
\begin{equation}
D^2+D^{*2}+2D^*D \geq - \frac 1{\sqrt{\alpha}}  D^*D - \frac {\sqrt{\alpha}}\pi \Lambda^2 \ .
\end{equation}
Together with $D^*D\leq \mbox{$\frac 2\pi$}\Lambda H_f$ and the {\it a
priori} estimate $\langle\Psi_0 |H_f|\Psi_0\rangle\leq \const \alpha$
this allows us to conclude that 
\begin{equation}
\langle\Psi_0 |\alpha
A^2|\Psi_0\rangle\geq \frac
\alpha\pi\Lambda^2-\const \alpha^{3/2} \ .
\end{equation}
By similar arguments, using
Schwarz' inequality as in (\ref{311}), with $a=1/\sqrt{\alpha}$, and
$\langle\Psi_0 |\,|d\Gamma(k)|^2\,|\Psi_0\rangle\leq \const \alpha$,
\begin{equation}
|\langle\Psi_0 |\sqrt{\alpha} d\Gamma(k)\cdot A |\Psi_0\rangle|\leq \const \alpha^{3/2} \ .
\end{equation}
We therefore obtain 
\begin{multline}\label{315}
\langle\Psi_0 |H_P|\Psi_0\rangle\geq |P|^2+\frac\alpha\pi \Lambda^2 -\const \alpha^{3/2}+ \\ \langle\Psi_0 |\, |d\Gamma(k)|^2-2P\cdot d\Gamma(k)+2\as P\cdot A+\as \vs\cdot B + H_f+\alpha^{3/2}|\Psi_0\rangle \ ,
\end{multline}
where we inserted the last term $\alpha^{3/2}$ because it will be
convenient later. The expression in the second line can be
rewritten as
\begin{multline}\label{rewr}
\sum_{n\geq 1}\Big[ \langle\psi_n|\, |d\Gamma(k)|^2 + H_f-2P\cdot d\Gamma(k)+\alpha^{3/2} | \psi_n\rangle \\+ 2\as \,\Re\,  \langle\psi_n| 2P\cdot D^*+\vs\cdot E^* | \psi_{n-1}\rangle\Big] \ ,
\end{multline}
where $\psi_n$ denotes the $n$-photon part of the wave function $\Psi_0$. 
We introduce the operator
\begin{equation}
\Q=|d\Gamma(k)|^2+H_f-2P\cdot d\Gamma(k)+\alpha^{3/2} \ ,
\end{equation}
which is positive for $|P|<\half$. 
By using Schwarz' inequality, we can estimate (\ref{rewr}) by
\begin{equation}\label{rewr3}
(\ref{rewr})\geq \sum_{n\geq 1} -\alpha \Big\langle\psi_{n-1}\Big|
(2P\cdot D+\vs\cdot E)\frac 1
{\Q}  (2P\cdot D^*+\vs\cdot E^*) \Big| \psi_{n-1}\Big\rangle \ . 
\end{equation}
We now investigate this term. The operator
$(2P\cdot D^*+\vs\cdot E^*)$ acting on $\psi_{n-1}$ creates a
photon. Explicitly, in momentum representation, suppressing the
polarization in the notation,
\begin{equation}
[E^* \psi_{n-1}](k_1,\dots k_n)=\frac 1{\sqrt n} \sum_{i=1}^n H(k_i)\psi_{n-1}(k_1,\dots,\not \!\! k_i,\dots, k_n) \ ,
\end{equation}
where $\not \!\! k_i$ means that $k_i$ is omitted from
consideration. An analogous expression holds for $D^*$. The right side
of (\ref{rewr3}) can therefore be split into two parts, one coming
from the \lq\lq diagonal terms\rq\rq , where both on the left and on
the right side the photon is created at the $i$'th position, and the
mixed terms. More precisely, denoting $F(k)= 2 P\cdot G(k)I_{\C^2}
+\vs\cdot H(k)$ and using permutation symmetry,
\begin{equation}
(\ref{rewr3})= -\alpha \sum_{n\geq 1} \big(I_n + II_n\big) \ ,
\end{equation}
where the two terms $I_n$ and $II_n$ are given by
\begin{equation}
I_n= \Big\langle F(k_n)\psi_{n-1}(k_1,\dots,k_{n-1}) \Big| \frac 1{\Q}\Big| F(k_n)\psi_{n-1}(k_1,\dots,k_{n-1})   \Big\rangle
\end{equation}
and
\begin{equation}
II_n=(n-1) \Big\langle F(k_n)\psi_{n-1}(k_1,\dots,k_{n-1}) \Big| \frac 1{\Q}\Big| F(k_1)\psi_{n-1}(k_2,\dots,k_{n})  \Big\rangle \ .
\end{equation}

To estimate $I_n$ from above, we write
$\Q=\A+b$, with ${\mathcal A}=|k_n|^2+|k_n|-2P\cdot k_n$, and use
\begin{equation}
\frac 1{{\mathcal A}+b}=\frac 1{{\mathcal A}}-\frac b{\A(\A+b)} 
\end{equation}
and the estimates $b\geq -2|k_n|\sum_{i=1}^{n-1}|k_i|$, $\A\geq |k_n|(1-2|P|)$ and $\A+b \geq |k_n|(1-2|P|)$ on the last term. Writing $\psi_{n-1}=\psi_{n-1}(k_1,\dots,k_{n-1})$ and 
\begin{equation}
W= W(k_1,\dots,k_n)= \frac 1{|k_n|^2+|k_n|-2P\cdot k_n}+\frac {2|k_n| \sum_{i=1}^{n-1} |k_i|}{|k_n|^2(1-2|P|)^2}
\end{equation}
for short,  we get
\begin{multline}
I_n\leq \Big\langle F(k_n)\psi_{n-1} \Big|W \Big| F(k_n)\psi_{n-1} \Big\rangle \\ =
4 \Big\langle P\cdot G(k_n)\psi_{n-1} \Big| W \Big| P\cdot G(k_n)\psi_{n-1}\Big\rangle \\+ \sum_{j=1}^3
\Big\langle H_j(k_n)\psi_{n-1} \Big|W \Big| H_j(k_n)\psi_{n-1}\Big\rangle \ , 
\end{multline}
where we used again the anti-commutation relations of the Pauli matrices and the fact that the mixed terms vanish, as explained in the upper bound. Inserting the expressions (\ref{defG}) and (\ref{defH}) for $G(k)$ and $H(k)$ yields (compare with (\ref{38})--(\ref{upbo}))
\begin{multline}\label{put1}
I_n\leq \|\psi_{n-1}\|^2 \frac 1{(2\pi)^2} 
\int_{|k|\leq \Lambda}  d^3k \frac { 4 \left(|P|^2|k|^2- |P\cdot k|^2\right) + 2|k|^4 } {|k|^3(|k|^2+|k|-2P\cdot k)}
\\+\langle \psi_{n-1}|H_f|\psi_{n-1}\rangle\frac 1{(2\pi)^2} 
\int_{|k|\leq \Lambda} d^3k  \frac{ 4 \left(|P|^2|k|^2- |P\cdot k|^2\right) + 2|k|^4
}{|k|^4(1-2|P|)^2}  \ .
\end{multline}

To estimate the term $II_n$ we use the crude estimate
\begin{multline}
|F(k_n)\psi_{n-1}|\leq \left( 2|P\cdot G(k_n)| + 3 |H(k_n)|\right) |\psi_{n-1}(k_1,\dots,k_{n-1})| \\ \equiv |F(k_n)|\, |\psi_{n-1}(k_1,\dots,k_{n-1})|  \ ,
\end{multline}
where the absolute values of $\psi_{n-1}$, $G$ and $H$ contain the
appropriate norms in $\C^{2}$. On the $n$-photon space $\Q\geq
(1-2|P|)(|k_1|+|k_n|+\alpha^{3/2})$, so we get the upper bound
\begin{multline}
II_n\leq  \frac{n-1}{1-2|P|} \int d^3 k_1\cdots d^3k_n \\ \times \frac { |\psi_{n-1}(k_1,\dots,k_{n-1})| |F(k_n)| |F(k_1)| |\psi_{n-1}(k_2,\dots,k_n)|}{ |k_1|+|k_n|+\alpha^{3/2} } \ .
\end{multline}
Introducing the photon density
\begin{equation}
\rho_{n-1}(k)=(n-1)\int d^3k_2 \cdots d^3k_{n-1} |\psi_{n-1}(k,k_2,\dots,k_{n-1})|^2
\end{equation}
and using Schwarz' inequality twice, we estimate
\begin{multline}\label{put2}
II_n\leq \frac 1{1-2|P|} \int d^3k_1 d^3k_n
\frac{\sqrt{\rho_{n-1}(k_1) \rho_{n-1}(k_n)} |F(k_1)||F(k_n)|}{
|k_1|+|k_n|+\alpha^{3/2} } \\ \leq \frac 1{1-2|P|} \langle\psi_{n-1}|H_f|\psi_{n-1}
\rangle\left(\int d^3k_1 d^3 k_n \frac{|F(k_1)|^2
|F(k_n)|^2}{|k_1||k_n|( |k_1|+|k_n|+\alpha^{3/2})^2}\right)^{\half}  .
\end{multline}
Since $|F(k)|\leq \const |k|^{-1/2}$ for small $k$ the last term diverges logarithmically as $\alpha\to 0$.   

Putting together (\ref{put1}) and (\ref{put2}) and using the {\it a priori} knowledge that $\langle\Psi_0|H_f|\Psi_0\rangle\leq\const \alpha$ we get 
\begin{multline}
\sum_{n\geq 1}\big( I_n+II_n\big) \leq   \frac 1{(2\pi)^2} 
\int_{|k|\leq \Lambda}  d^3k \frac { 4 \left(|P|^2|k|^2- |P\cdot k|^2\right) + 2|k|^4 } {|k|^3(|k|^2+|k|-2P\cdot k)}
\\+ O( \alpha \ln(1/\alpha)) \ .
\end{multline}
We insert this into (\ref{315}) and arrive at the  lower bound
\begin{multline}
\E_P\geq |P|^2 + \frac \alpha\pi \Lambda^2 - \frac\alpha{(2\pi)^2}  
\int_{|k|\leq \Lambda}  d^3k \frac { 4 \left(|P|^2|k|^2- |P\cdot k|^2\right) + 2|k|^4 } {|k|^3(|k|^2+|k|-2P\cdot k)}
\\ - O(\alpha^{3/2}) \ .
\end{multline}
Together with the upper bound (\ref{upbo}) this proves (\ref{ep}), since 
\begin{multline}
2|k|\left(\frac 1{|k|^2+|k|-2P\cdot k}-\frac 1{|k|^2+|k|}\right)\\ = \frac {2|k|}{(|k|^2+|k|)^2} \left(2 P\cdot k + \frac {|2 P\cdot k|^2}{|k|^2+|k|-2P\cdot k}\right) \ , 
\end{multline}
and the first term on the right side vanishes after integration over $k$. 
\end{proof}

The coefficient of $|P|^2$ in an expansion of $\E_P$ around $P=0$ determines the physical mass $m$ as a function of $m_0$, $\alpha$ and $\Lambda$. Since
\begin{multline}
\int_{|k|\leq \frac \Lambda{2m_0}}
d^3k \frac{|k|}{|k|^2+|k|} \left(\frac {|P|^2}{|k|^2} + |P\cdot k|^2 \left(\frac 2{(|k|^2+|k|)^2} - \frac 1{|k|^4}\right)\right) 
= \\
|P|^2 \frac {16\pi} 3\left[ \ln\left(1+\frac \Lambda{2m_0}\right)-\frac 34 \frac{\Lambda(\Lambda+\mbox{$\frac 43$}m_0)}{(\Lambda+ 2m_0)^2}\right] \ ,
\end{multline}
Equation~(\ref{ep}) leads to the following relation of $m$ and $m_0$,
to leading order in $\alpha$:
\begin{equation}\label{defph}
m=m_0 \left(1+\alpha \frac {16} {3\pi}\left[ \ln\left(1+\frac \Lambda{2m_0}\right) -\frac 34 \frac{\Lambda(\Lambda+\mbox{$\frac 43$}m_0)}{(\Lambda+2m_0)^2}\right]\right)  \ .
\end{equation}
This relation is decisive when renormalizing the bare mass in the
expression for the binding energy derived in the next section. Note
that for any given positive values of $m$, $\alpha$ and $\Lambda$
there is a unique positive solution $m_0$ to the equation above. This is 
in
contrast to the analogous expression in dipole approximation, where
the term in square brackets is replaced by a factor linear in
$\Lambda/m_0$, and consequently a positive solution only exists for
$\alpha\Lambda$ small enough. More precisely, in dipole approximation
the physical mass can be calculated explicitly, not only to leading
order in $\alpha$, and the result is
\begin{equation}
m=m_0+\frac {4\alpha}{3\pi} \Lambda \ .
\end{equation}
This was first shown by van Kampen in his thesis, as described in Kramers' biography \cite[Sect.~16.III.D]{kramers}. 

\section{Binding energy}\label{bindsect}

In this section we shall calculate the binding energy of an electron in the
field of nucleus of charge $Z>0$. As we explained in the Introduction,
we want to consider the case of small $\alpha$ with $Z\alpha$ not
necessarily small. To avoid confusion, we denote the fine
structure constant appearing in the external potential by
$\beta$. I.e., we add to the Hamiltonian the potential 
\begin{equation}\label{pot}
V(x)=-\frac{\beta Z}{|x|} \ ,
\end{equation}
with $\beta\approx 1/137$. More general external potentials can be
treated in the same manner, but for the sake of simplicity we shall
not do so here.

Let $e_0=-\half m_0 (\beta Z)^2$ be the ground state energy of the
hydrogen atom without coupling to the photon field, and denote the
corresponding ground state wave function by $\phi_0$.
Let $\E(V)$ denote the ground state energy of (\ref{rpf}). The existence of a ground state for $H$ has recently been shown in \cite{GLL}.
 
The following Theorem gives $\E(V)$ to leading order in $\alpha$. To obtain the binding energy, one has to subtract it from the self energy $\E(0)$, which will be done in Theorem~\ref{bindthm}.
We introduce the notation that $A(x)=D(x)+D^*(x)$ and $B(x)=E(x)+E^*(x)$, where $D$ and $E$ denote the part of $A$ and $B$, respectively, containing only annihilation operators. 

\begin{thm}[{\bf Ground State Energy}]\label{uppevthm}
 With $\A$ denoting the operator $\A=\mbox{$\frac 1{2m_0}$}|p|^2+V-e_0 + H_f$, the ground state energy of (\ref{rpf}) with external potential (\ref{pot}) is given by
\begin{equation}\label{uppev}
\E(V)= e_0 +\frac \alpha{2m_0\pi} \Lambda^2  - \frac {\alpha}{4m_0^2}\Big\langle\phi_0 \Big| 4 p\cdot D \frac 1\A p\cdot D^* + E\cdot \frac 1\A  E^* \Big| \phi_0\Big\rangle  + O(\alpha^{3/2})  .
\end{equation}
\end{thm}

\begin{proof}
We set again $m_0=\half$ for simplicity, and start with the {\it upper bound:} We use as a trial state 
\begin{equation}
\Psi=\left\{\phi_0 \uparrow, \frac {-\as}{|p|^2+V-e_0+|k|}\Big[2 p\cdot D^*+ \vs\cdot E^* \Big]\phi_0 \uparrow ,0,\dots\right\}  ,
\end{equation}
with $E^*$, $D^*$ and $\phi_0$ as explained above. Note that in contrast to (\ref{3.5}) there is no need for an infrared cut-off, since the norm of $\Psi$ is finite. This can be seen as follows. An infrared problem can only arise from the overlap of $p\cdot D^* \phi_0$ with $\phi_0$, since $|p|^2+V-e_0$ has a spectral gap above zero. Since $p\phi_0$ is orthogonal to $\phi_0$, we can estimate the projection of $p\cdot D^* \phi_0$ onto $\phi_0$ (for fixed $k$) by
\begin{multline}
  |\langle \phi_0 | p\cdot D^* \phi_0\rangle| = |G(k)\cdot \langle \phi_0 | e^{-ik\cdot x} p \phi_0\rangle | \\ = |G(k)\cdot  \langle \phi_0 | (e^{-ik\cdot x}-1)  p \phi_0\rangle| \leq  |G(k)|\, |k| \, \|p\phi_0\|^2 \|x\phi_0\|^2 \ ,
\end{multline}
where we used that $|e^{-i k\cdot x}-1|\leq  |k| |x|$. The thus obtained additional factor $|k|$ makes the integral over $k$ finite.

Similarly to the
calculation in the previous section, we obtain
\begin{multline}
\langle \Psi|H|\Psi\rangle = \left(e_0 + \frac \alpha\pi \Lambda^2\right) \langle \Psi|\Psi\rangle \\ - \alpha\Big\langle (2p\cdot D^*+\vs \cdot E^*)\phi_0\uparrow \Big| \frac 1{\A} \Big|  (2p\cdot D^*+\vs \cdot E^*)\phi_0 \uparrow \Big\rangle  + O(\alpha^2) \  .
\end{multline}
Using the anti-commutation relations for $\sigma$, we see that the mixed terms vanish, and we arrive at (\ref{uppev}) as an upper bound. More precisely,
\begin{equation}\label{45}
\Re\, \Big\langle p\cdot D^* \phi_0 \uparrow \Big| \frac 1\A\Big| \vs\cdot E^* \phi_0 \uparrow \Big\rangle = 0
\end{equation}
and
\begin{equation}\label{46}
\Im\, \Big\langle E^*_i \phi_0 \Big| \frac 1\A\Big|  E^*_j \phi_0 \Big\rangle = 0 \ ,
\end{equation}
as can be seen by using the fact the $G$ is real and $H$ is purely
imaginary, and that the operator $\A$ commutes with the reflection
$x\to -\x$.

\noindent 
{\it Lower bound:} Proceeding as in the proof of the lower bound in
Theorem~\ref{T1} (c.f. also \cite{H1}), it is easy to see that, in the
ground state $\Psi_0$, $\langle \Psi_0|H_f|\Psi_0\rangle\leq \const
\alpha$. Moreover,
\begin{multline}
\langle\Psi_0 |H|\Psi_0\rangle\geq e_0+ \frac \alpha\pi \Lambda^2 -\const \alpha^{3/2} \\+ \langle\Psi_0 |\, |p|^2 + V - e_0 +2\as p\cdot A+\as \vs\cdot B + H_f|\Psi_0\rangle 
\end{multline}
(compare with (\ref{315})). 
The expression in the second line can be rewritten as
\begin{multline}\label{rewr2}
\langle\psi_0| \, |p|^2+V-e_0| \psi_0\rangle +
\sum_{n\geq 1} \langle\psi_n|\,  |p|^2+V-e_0 + H_f | \psi_n\rangle \\+ 2\as\, \Re\,  \langle\psi_n| 2p\cdot D^*+\vs\cdot E^* | \psi_{n-1}\rangle \ .
\end{multline}
With $\A$ as stated in the theorem and $F^*=2p\cdot D^*+\vs\cdot E^*$, we get as a lower bound for (\ref{rewr2}), using Schwarz' inequality, 
\begin{equation}\label{lowerb}
\langle \psi_0|  \A | \psi_0\rangle - \alpha \sum_{n\geq 0}
\Big\langle\psi_{n}\Big| F\cdot \frac
1{\A} F^* \Big| \psi_{n}\Big\rangle \ .
\end{equation}
To bound the terms with $n\geq 1$ from below, we need the following Lemma. 

\begin{lem} Let $\Nb=d\Gamma(1)$ denote the number operator. Then
\begin{equation}
F \cdot \frac 1{\A} F^* \leq \const (1+\Nb) \ .
\end{equation}
\end{lem}

\begin{proof}
By Schwarz' inequality
\begin{equation}
F \cdot \frac 1{\A} F^* \leq 2 \vs\cdot E \frac 1\A \vs\cdot E^* + 8  D\cdot p \frac 1\A p\cdot D^*\ .
\end{equation}
We first consider the last term. Let $p_i$, $i=1,2,3$, denote the components of $p$. We claim that 
\begin{equation}
p_i\frac 1{|p|^2+V-e_0+H_f}p_i\leq 1 + \frac {|e_0|}{H_f} \ .
\end{equation}
It suffices to prove this for $H_f$ some positive number $\mu$. Since
\begin{equation}
\eps p_i^2\leq \eps |p|^2 \leq |p|^2+V- e_0/(1-\eps)
\end{equation}
for $0\leq \eps < 1$, we get
\begin{equation}
 p_i\frac 1{|p|^2+V-e_0/(1-\eps)}p_i\leq \frac 1\eps \ . 
\end{equation}
With $\eps = \mu/(\mu + |e_0|)$ this gives
\begin{equation}
p_i\frac 1{|p|^2+V-e_0+\mu}p_i\leq 1 + \frac {|e_0|}{\mu} \ .
\end{equation}
Using that $\A\geq H_f$ and again Schwarz' inequality for the terms with $p_i$ and $p_j$ for $i\neq j$, we thus obtain
\begin{equation}\label{416}
F \cdot \frac 1\A F^* \leq 2 \vs\cdot E \frac 1{H_f} \vs\cdot E^* + 24  D\cdot\left(1+ \frac {|e_0|}{H_f}\right) D^* \ .
\end{equation}
Consider now the operator $ D_i\left(1+ \frac {|e_0|}{H_f}\right)
D_i^* $ restricted to the $(n-1)$-photon sector. By use of Schwarz'
inequality we can estimate
\begin{multline}
\Big\langle \psi_n \Big|   D_i\left(1+ \frac {|e_0|}{H_f}\right) D_i^* \Big|\psi_n\Big\rangle \\ \leq n \Big\langle \psi_{n-1} e^{ik_n\cdot x} G_i(k_{n}) \Big| 1+ \frac {|e_0|}{H_f}\Big|  \psi_{n-1} e^{ik_n\cdot x} G_i(k_{n}) \Big\rangle \ ,
\end{multline} 
where $\psi_{n-1}$ is short for
$\psi_{n-1}(x,k_1,\dots,k_{n-1})$. Since $H_f\geq |k_n|$ in the
\mbox{$n$-photon} sector, the last expression is bounded above by
\begin{equation}
n \|\psi_n\|^2 \sum_{\lambda=1}^2 \int d^3k |G_i^\lambda(k)|^2 \left(1+ \frac
{|e_0|}{|k|}\right) \ .
\end{equation}
Proceeding analogously for the other terms in (\ref{416}) we prove the desired result.
\end{proof}

From \cite[Thm.~6.1]{GLL} we infer that in the ground state $\Psi_0$ we have 
$\langle\Psi_0|\Nb|\Psi_0\rangle \leq \const \alpha$, where the constant depends only on $\Lambda$. Using this we conclude from the Lemma above that  
\begin{equation}\label{lowerbb}
(\ref{lowerb})\geq  \langle \psi_0|  \A | \psi_0\rangle - \alpha 
\Big\langle\psi_0\Big| F\cdot  \frac
1{\A} F^* \Big| \psi_0\Big\rangle - \const \alpha^2 \ ,
\end{equation}
and also that $\|\psi_0\|^2 \geq 1- \const \alpha $. It therefore follows from first order perturbation theory \cite{kato} that, for small enough $\alpha$, 
\begin{equation}
(\ref{lowerbb}) \geq   - \alpha 
\Big\langle\phi_0 \uparrow \Big| F\cdot \frac
1{\A} F^* \Big| \phi_0 \uparrow\Big\rangle - \const \alpha^2 \ .
\end{equation}
The spin $\uparrow$ is just one possible choice, any other spin will do. 
As already noted in the upper bound (c.f. (\ref{45}) and (\ref{46})),
\begin{equation}
\Big\langle\phi_0\uparrow\Big| F\cdot \frac
1{\A} F^* \Big| \phi_0\uparrow\Big\rangle = 
\Big\langle \phi_0\Big | 4 p\cdot D \frac 1\A p\cdot D^* +  E\cdot \frac 1\A E^* \Big|\phi_0\Big \rangle  \ , 
\end{equation}
which finishes the proof of the lower bound.
\end{proof}

With the expression for the ground state energy $\E(V)$ in hand, we
can subtract it from the self energy to obtain the binding
energy. This generalizes a result in \cite{H1}, where a lower bound on
the binding energy was obtained. Moreover, the self energy was
calculated, to leading order in $\alpha$, in \cite{H1}, and we will
use the result from there. Note that both the ground state energy and
the self energy are, to order $\alpha$, linearly divergent in
$\Lambda$ (for fixed $m_0$). The binding energy only shown a
logarithmic divergence, which will be removed by mass renormalization
in the next section.

\begin{thm}[{\bf Binding Energy}]\label{bindthm}
With $\B$ denoting the operator $\B=\mbox{$\frac 1{2m_0}$}|p-k|^2 + V
- e_0 + |k|$, the binding energy of (\ref{rpf}) with external
potential (\ref{pot}) is given by
\begin{multline}\label{binden}
\E(0)-\E(V)=-e_0  +  \frac{\alpha}{m_0^2}\sum_{i,j=1}^3
\Big\langle p_i\phi_0\otimes G_i \Big|  \frac 1{\B}  \Big| p_j\phi_0\otimes G_j \Big\rangle \\ + \frac{\alpha}{m_0^2} \sum_{i=1}^3 \Big\langle \phi_0\otimes H_i \Big| \frac {k\cdot p}{|k|^2+2m_0|k|}\, \frac 1{\B} \, \frac{ k\cdot p}{|k|^2+2m_0|k|}  \Big| \phi_0\otimes H_i \Big\rangle + O( \alpha^{3/2}) \ .
\end{multline}
\end{thm}

\begin{proof} We set again $m_0=\half$ for simplicity. 
From \cite[Thm.~1]{H1} we know that
\begin{eqnarray}\nonumber
\E(0)&=&\frac\alpha\pi \Lambda^2 -\alpha \Big\langle 0\Big| E(0)\cdot \frac 1{|k|^2+|k|} E^*(0) \Big|0\Big\rangle +O(\alpha^2) \\ \label{know} &=& \frac {2\alpha}{\pi} \big(\Lambda - \ln(1+\Lambda)\big) + O(\alpha^2) \ ,
\end{eqnarray}
where $|0\rangle$ denotes the vacuum in $\F_b$. We claim that the vacuum expectation value in the first line of (\ref{know}) can be written as 
\begin{equation}
-\alpha \Big\langle\phi_0\Big| E\cdot \frac 1{|p+k|^2+V-e_0 + |k|^2 + |k| } E^* \Big|\phi_0 \Big\rangle \ ,
\end{equation}
where $E=E(x)$. To see this, consider the unitary transformation $U$
on the one-photon space $L^2(\R^3,d^3 x; \C^2)\otimes L^2(\R^3,d^3 k;
\C^2)$ given by
\begin{equation}\label{uni}
(U\psi)(x,k)=e^{ik\cdot x} \psi(x,k) \ ,
\end{equation}
suppressing the dependence on spin and polarization in the notation. 
We have
\begin{equation}
(U E_i^* \phi_0)(x,k) = H_i(k) \phi_0(x) = \phi_0 \otimes H_i  \ .
\end{equation}
Moreover, 
\begin{multline}
U \Big( |p+k|^2 + V(x) - e_0 + |k|^2 + |k| \Big) U^* =  |p|^2 + V(x) - e_0 + |k|^2 + |k|  \\ = \Big( |p|^2 + V - e_0 \Big)\otimes 1  +  1 \otimes \Big( |k|^2 + |k| \Big) \ , 
\end{multline}
which proves the claim. 

Introducing the notation $\B'= |p|^2 + V - e_0 + |k|^2 + |k|=\B+2p\cdot k$ and using this unitary transformation, we have
\begin{multline}
\Big\langle\phi_0 \Big| E\cdot \frac 1{\B}  E^* \Big| \phi_0\Big\rangle - \Big\langle\phi_0\Big| E\cdot \frac 1{|p+k|^2+V-e_0 + |k|^2 + |k| } E^* \Big|\phi_0 \Big\rangle
 \\ = \sum_{i=1}^3 \Big\langle \phi_0\otimes H_i \Big| \frac{1}{\B} -  \frac{1}{\B'}\Big| \phi_0\otimes H_i \Big\rangle\\  =\sum_{i=1}^3 \Big\langle \phi_0\otimes H_i \Big| \frac 1{\B'} 2 p\cdot k \frac 1{\B'} + \frac 1{\B'} 2p\cdot k \frac 1{\B} 2 p \cdot k \frac 1{\B'}  \Big| \phi_0\otimes H_i \Big\rangle \ .
\end{multline}
The first term in the last expression is zero by symmetry with respect
to the reflection $x\to-\x$. Moreover, ${\B'}^{-1}\phi_0\otimes
H_i=\phi_0\otimes H_i/(|k|^2+|k|)$. Hence, using Theorem \ref{uppevthm}
and (\ref{know}), and the unitary transformation (\ref{uni}), we obtain, up to an error not bigger that order
$\alpha^{3/2}$,
\begin{multline}
\E(0)-\E(V)=-e_0 + 4\alpha \sum_{i,j=1}^3
\Big\langle p_i\phi_0\otimes G_i \Big|  \frac 1{\B}  \Big| p_j\phi_0\otimes G_j \Big\rangle \\  + \alpha\sum_{i=1}^3 \Big\langle \phi_0\otimes H_i \Big| \frac {2k\cdot p}{|k|^2+|k|}\, \frac 1{\B} \, \frac{2 k\cdot p}{|k|^2+|k|}  \Big| \phi_0\otimes H_i \Big\rangle  \ ,
\end{multline}
proving the Theorem.
\end{proof}

\section{Renormalization}\label{rensect}

In the previous section, we calculated the binding energy to leading
order in $\alpha$. However, the bare mass enters this calculation,
which is not a physical quantity. We have to replace it by the
physical mass, which was, to leading order in $\alpha$, calculated in
Section \ref{massrensect}. Note that both expressions show a
logarithmic dependence on $\Lambda$. It turns out the these
divergences cancel and a finite result will be obtained.

In the two terms in (\ref{binden}) of order
$\alpha$ we can simply replace $m_0$ by $m$ since we are only
interested in a leading order calculation. This also affects $e_0$ and
$\phi_0$; i.e., from now on $e_0=-\half m (\beta Z)^2$ and $\phi_0$ is
the ground state wave function of $\mbox{$\frac 1{2m}|p|^2 + V$}$.

The $m_0$ in the first term $e_0$ in (\ref{binden}) has to be replaced
by the expression (\ref{defph}), however. This leads to
\begin{multline}\label{renbindd2}
\E(0)-\E(V)= \frac m2 (\beta Z)^2\left(1-\alpha \frac {16} {3\pi}\left[ \ln\left(1+\frac \Lambda{2m}\right) -\frac 34 \frac{\Lambda(\Lambda+\mbox{$\frac 43$}m)}{(\Lambda+2m)^2}\right]\right)  \\  
+  \frac{\alpha}{m^2}\sum_{i,j=1}^3
\Big\langle p_i\phi_0\otimes G_i \Big|  \frac 1{\B}  \Big| p_j\phi_0\otimes G_j \Big\rangle \\
+ \frac{\alpha}{m^2} \sum_{i=1}^3 \Big\langle \phi_0\otimes H_i \Big| \frac {k\cdot p}{|k|^2+2m|k|}\, \frac 1{\B} \, \frac{ k\cdot p}{|k|^2+2m|k|}  \Big| \phi_0\otimes H_i \Big\rangle  \ ,
\end{multline}
with $\B=\mbox{$\frac 1{2m}$}|p-k|^2 + V - e_0 + |k|$. We suppress
higher order terms in $\alpha$ from now on.

We shall show that (\ref{renbindd2}) has a finite limit
as $\Lambda\to\infty$, i.e., the cut-off can be removed. Let $\B'$ denote the operator $\B'=\mbox{$\frac 1{2m}$}|p|^2 + V - e_0 + \mbox{$\frac 1{2m}$}|k|^2+ |k|=\B+ p\cdot k/m$. An easy calculation, using (\ref{defG}) and (\ref{defH}) and the fact that 
\begin{equation}
\sum_\lambda G^\lambda_i(k)G^\lambda_j(k)= \frac1{(2\pi)^2|k|^3}\left(|k|^2\delta_{ij}-k_i k_j\right) \ ,
\end{equation}
gives
\begin{multline} 
\sum_{i,j=1}^3
\Big\langle p_i\phi_0\otimes G_i \Big|  \frac 1{\B'}  \Big| p_j\phi_0\otimes G_j \Big\rangle \\ +
\sum_{i=1}^3 \Big\langle \phi_0\otimes H_i \Big| \frac {k\cdot p}{|k|^2+2m|k|}\, \frac 1{\B'} \, \frac{ k\cdot p}{|k|^2+2m|k|}  \Big| \phi_0\otimes H_i \Big\rangle \\=m\sum_{j=1}^3 \Big\langle p_j \phi_0 \Big| f\left(\frac  {\mbox{$\frac 1{2m}$}|p|^2+V-e_0}{2m}, \frac{\Lambda}{2m} \right) \Big| p_j\phi_0 \Big\rangle \ .
\end{multline}
Here $f$ is the function
\begin{equation}
f(e,\Lambda)=\frac{4}{3\pi} \int_0^{\Lambda} dk
\frac{k^5}{e+k^2+k}\left(\frac 1{k^4}+\frac1{(k^2+k)^2}\right) \ .
\end{equation}
Note that $\sum_j \langle p_j\phi_0|p_j\phi_0\rangle =
2m|e_0|$, and that
\begin{equation}
f(0,\Lambda)= \frac{8}{3\pi}\left(\ln(1+\Lambda)-\frac 34 \frac{\Lambda(\Lambda+\mbox{$\frac 23$})}{(\Lambda+1)^2}\right) \ .
\end{equation}
Therefore (\ref{renbindd2}) can be written as
\begin{multline}\label{renbindd3}
\E(0)-\E(V)= \frac m2 (\beta Z)^2 \\ +\frac\alpha\mph 
\sum_{j=1}^3 \Big\langle p_j \phi_0 \Big| f\left(\frac  {\mbox{$\frac 1{2m}$}|p|^2+V-e_0}{2m}, \frac{\Lambda}{2m} \right)-f\left(0,\frac\Lambda{2m}\right) \Big| p_j\phi_0 \Big\rangle \\ 
+  \frac{\alpha}{m^2}\sum_{i,j=1}^3
\Big\langle p_i\phi_0\otimes G_i \Big|  \left(\frac 1{\B}-\frac 1{\B'}\right)  \Big| p_j\phi_0\otimes G_j \Big\rangle
\\+ \frac{\alpha}{m^2} \sum_{i=1}^3 \Big\langle \phi_0\otimes H_i \Big| \frac {k\cdot p}{|k|^2+2m|k|}\,\left( \frac 1{\B}-\frac 1{\B'}\right) \, \frac{ k\cdot p}{|k|^2+2m|k|}  \Big| \phi_0\otimes H_i \Big\rangle \ .
\end{multline}

By definition $\B=\B'-b$ with $b=p\cdot k/m$. However, it is easy to see that the expressions (\ref{renbindd2}) and (\ref{renbindd3}) do not change if we replace $\B$ by $\B'+b$. Therefore we can replace $1/\B$ by $\half(1/(\B'-b)+1/(\B'+b))$, and instead of $1/\B-1/\B'$ in (\ref{renbindd3}) we can write
\begin{equation}
\frac 12\left(\frac 1{\B'-b}+\frac 1{\B'+b}\right)-\frac 1{\B'} =  \frac 1{\B'}b\frac1{\B'-b{\B'}^{-1}b}b\frac 1{\B'}  \ .
\end{equation}
For given operators $h$ and $p$ let $C_k(h,p)$ denote the operator
\begin{multline}
C_k(h,p)= \frac 1{h+|k|^2+|k|}\, p\cdot k \times \\ \times \frac 1{h+|k|^2+|k| - p\cdot k (h+|k|^2+|k|)^{-1} p\cdot k} \, p\cdot k \, \frac 1{h+|k|^2+|k|} \ .
\end{multline}
We define the operator valued matrix $\T^\Lambda$ by its components 
\begin{multline}\label{deftl}
\T_{ij}^\Lambda(h,p)=\\ \frac 1{2\pi^2} \int_{|k|\leq \Lambda} d^3k 
\left(\frac{2 |k| k_i k_j}{(|k|^2+|k|)^2}+\frac 1{|k|^3}\left(|k|^2 \delta_{ij}-k_i k_j\right)\right) C_k(h,p) \ ,
\end{multline}
being an operator on $L^2(\R^3,d^3x;\C^2)\otimes \C^3$. 
Using again (\ref{defG}), (\ref{defH}) and the orthogonality relations of $\eps_\lambda(k)$ and $k$, we can rewrite (\ref{renbindd3}) as 
\begin{multline}\label{renbindd4}
\E(0)-\E(V)= \frac m2 (\beta Z)^2 \\ +\frac\alpha\mph 
\sum_{j=1}^3 \Big\langle p_j \phi_0 \Big| f\left(\frac  {\mbox{$\frac 1{2m}$}|p|^2+V-e_0}{2m}, \frac{\Lambda}{2m} \right)-f\left(0,\frac\Lambda{2m}\right) \Big| p_j\phi_0 \Big\rangle \\ 
+ \frac{\alpha}{m} \sum_{i,j=1}^3 \Big\langle p_i\phi_0 \Big|\, \T_{ij}^{\Lambda/(2m)}\left(\frac  {\mbox{$\frac 1{2m}$}|p|^2+V-e_0}{2m}, \frac p m \right) \Big| p_j\phi_0\Big\rangle  \ .
\end{multline}
Both terms on the right side have a nice limit as $\Lambda\to\infty$. We define
\begin{multline}\label{defg}
\Ss(e)= \lim_{\Lambda\to\infty} f(0,\Lambda)-f(e,\Lambda) \\ = \frac{4}{3\pi} \int_0^{\infty} dk
\frac{e\, k^5}{(k^2+k)(e+k^2+k)}\left(\frac 1{k^4}+\frac1{(k^2+k)^2}\right) 
\end{multline}
and $\T(h,p)$ by 
\begin{equation}\label{defT}
\T_{ij}(h,p)=\T_{ij}^\infty(h,p) \ .
\end{equation}
Note that this operator, for $h=\mbox{$\frac 1{2m}$}|p|^2+V-e_0$, is well defined. To see this,
we estimate
\begin{equation}
\frac 1{\B'-b{\B'}^{-1}b}\leq \frac 1{|k|}
\end{equation}
and 
\begin{equation}
\frac1{m^2}(p\cdot k)^2 \leq \frac1{m^2} |p|^2 |k|^2 \leq |k|^2 \frac2{\eps m} \left(
\mbox{$\frac 1{2m}$}|p|^2+V-e_0/(1-\eps)\right) 
\end{equation}
for all $0<\eps<1$. 
Since $\B'\geq |k|^2/(2m)+|k|$ we get 
\begin{equation}\label{tbound}
\T\left(\frac  {\mbox{$\frac 1{2m}$}|p|^2+V-e_0}{2m}, \frac p m \right)\leq  \frac2{\eps m} I_{\C^3} \left(
\mbox{$\frac 1{2m}$}|p|^2+V-e_0/(1-\eps)\right)  \frac{16}{9\pi} 
\end{equation}
as an operator on $L^2(\R^3,d\x;\C^2)\otimes \C^3$. 
The last factor comes from the integral
\begin{equation}
\frac 1{3\pi^2} \int_{\R^3} d^3k \frac{|k|^4}{(|k|^2+|k|)^2} \left(\frac 1{|k|^4}+\frac 1{(|k|^2+|k|)^2}\right) = \frac{16}{9\pi} \ .
\end{equation}

We see that, for $h=\mbox{$\frac 1{2m}$}|p|^2+V-e_0$, both the operators $\T^{\Lambda}(h/2m,p/m)$ and $f(0,\Lambda)-f(h/2m,\Lambda)$ are monotone increasing in $\Lambda$, and bounded by $\const (1+h)$ independent of $\Lambda$. Passing to the limit $\Lambda\to\infty$, we have thus proved:

\begin{thm}[{\bf Renormalized Binding Energy}]\label{renbindthm}
To leading order in $\alpha$, the renormalized binding energy, after removing the cut-off, is given by
\begin{multline}\label{renbind}
\E(0)-\E(V)= \frac m2 (\beta Z)^2  -\frac\alpha\mph 
\sum_{j=1}^3 \Big\langle p_j \phi_0 \Big|\, \Ss\left(\frac  {\mbox{$\frac 1{2m}$}|p|^2+V-e_0}{2m} \right) \Big| p_j\phi_0 \Big\rangle \\ 
+ \frac{\alpha}{m} \sum_{i,j=1}^3 \Big\langle p_i\phi_0 \Big|\, \T_{ij}\left(\frac  {\mbox{$\frac 1{2m}$}|p|^2+V-e_0}{2m}, \frac p m \right) \Big| p_j\phi_0\Big\rangle  \ ,
\end{multline}
with $\Ss$ and $\T$ the positive operators defined in (\ref{defg}), (\ref{defT}) and (\ref{deftl}).
\end{thm}

Note that, by scaling, the right side of (\ref{renbind}) is $m$ times a function of $\beta Z$. This is of course clear from the viewpoint of physical dimensions, since, after removing $\Lambda$, $m$ is the only energy scale in the problem.  

The function $\Ss$ is monotone increasing and concave, with
\begin{equation}\label{gsmall}
\Ss(e)\approx \frac {4}{3\pi} e \ln (1/e) \quad {\rm for\ } e\ll 1 \ ,
\end{equation}
and
\begin{equation}
\Ss(e)\approx \frac {4}{3\pi} \ln (e) \quad {\rm for\ } e\gg 1 \ .
\end{equation}
Because of the logarithmic factor appearing in (\ref{gsmall}), the
right side of (\ref{tbound}) is, for $e_0\ll m$, much smaller than the
second term in (\ref{renbind}). I.e., for $Z\ll 1/\beta\approx 137$, the
term with~$\T$  is negligible compared to the term with~$\Ss$.

Note that if we neglect $\T$ and approximate $\Ss(e)$ by $\mbox{$\frac
4{3\pi}$} e \ln(1/e)$ we obtain exactly Bethe's result for the ground state energy shift 
\cite{Be}. For this result it is not necessary to include the $B$
field in the Hamiltonian, since the logarithmic factor in $\Ss(e)$
for small $e$ is entirely due to the term $1/|k|^4$ in (\ref{defg}),
which comes from the $p\cdot A$ term in the Hamiltonian. The part with
$1/(|k|^2+|k|)^2$, which stems from the $\vs
\cdot B$ term, is, when divided by $e$, bounded as $e\to 0$.

For $\beta Z$ not too small, there is a significant difference between
our expression (\ref{renbind}) and Bethe's formula. First of all,
there is a different behavior of $\Ss(e)$ for large $e$, and secondly
the term $\T$ contributes, with a different sign than $\Ss$. It would
be interesting to evaluate the terms in (\ref{renbind}) numerically,
given the physical values of $\alpha=\beta$ and $Z$. For small $Z$,
the result will essentially agree with Bethe's expression, whereas for
larger values of $Z$ our formula should be closer to the
experimentally observed value.

\bigskip
\noindent {\bf Remark.} 
The appearance of the second term $\T$ in Theorem~\ref{renbindthm} is
essentially due to the fact that the physical mass was obtained at
total momentum $P=0$, whereas the ground state of an electron in
the field of a nucleus shows a definite momentum distribution. In
fact, a definition of the physical mass of the state $\phi_0$ by the
relation
\begin{equation}
\langle \phi_0 | \E_p-\E_0 |\phi_0\rangle = \frac 1{2m} \langle \phi_0 |\, |p|^2 |\phi_0\rangle \ ,
\end{equation}
where $p=-i\nabla_x$ is the electron momentum operator, would account
for a subtraction of a term similar to $\T$. From a physical point of view,
however, this definition is not very satisfactory since one wants to
compare the binding energy with the rest mass of the electron and not
with a mass defined in dependence on the state of the system.
\bigskip

To see the order of magnitude of the shift of the binding energy due to the presence of the quantized radiation field to leading order in $\alpha$, 
we now present a rough lower bound. The last term in (\ref{renbind}) is
positive and can be neglected. For the first term, we use concavity of
$\Ss$ and Jensen's inequality to get
\begin{multline}\label{renbind2}
\E(0)-\E(V) \geq  \frac{\mph}2 (\beta Z)^2 \\- \frac{\alpha}{\mph} \sum_{j=1}^3 \langle p_j\phi_0|p_j \phi_0\rangle \,\Ss\left(\frac 1{2\mph}\frac{ \sum_{j=1}^3 \Big\langle p_j\phi_0\Big|\mbox{$\frac 1{2m}$}|p|^2+V-e_0\Big|p_j \phi_0\Big\rangle}{\sum_{j=1}^3 \langle p_j\phi_0|p_j \phi_0\rangle}\right) \ . 
\end{multline}
Using that $\sum_{j=1}^3 \langle p_j\phi_0|p_j \phi_0\rangle=2m |e_0|$ and 
\begin{multline}
 \sum_{j=1}^3 \Big\langle p_j\phi_0\Big|\mbox{$\frac 1{2m}$}|p|^2+V-e_0\Big|p_j \phi_0\Big\rangle \\ = -\frac 12 \sum_{j=1}^3 \Big\langle\phi_0\Big|\Big [ p_j, \Big[p_j, \mbox{$\frac 1{2m}$}|p|^2+V-e_0
\Big]\Big] \Big| \phi_0\Big\rangle \\ =  2\pi (\beta Z) |\phi_0(0)|^2=2\mph^3 (\beta Z)^4 \ , 
\end{multline}
we get
\begin{equation}\label{renbind3}
\E(0)-\E(V) \geq  \frac{\mph}2 (\beta Z)^2 - \mph \alpha (\beta Z)^2 \, \Ss\big((\beta Z)^2\big) \ .
\end{equation}
In nature $\alpha=\beta$, and therefore
\begin{equation}
\E(0)-\E(V) \geq  \frac{\mph}2 (\beta Z)^2\Big[1-2\beta \, \Ss\big((\beta Z)^2\big)\Big] \ .
\end{equation}
Inserting $\beta=1/137$ and $Z=1$ for the hydrogen atom gives a lower
bound on the shift of the binding energy of $-12.09 *10^3$ MHz, which
is off the experimental value by a factor of about $1.5$ (see, e.g.,
the appendix of \cite{kino} for a detailed discussion). This is of
course only a rough lower bound, the true value given by
(\ref{renbind}) is probably much closer. It should be noted, however,
that the shift considered here is only the one due to
the quantized radiation field, which is smaller than the shift due to
relativistic effects.

\section{Lamb Shift}\label{lambsect}

Due to the presence of the quantized photon field, the ground state is
the only eigenstate of the Hamiltonian \cite{BFS}. All the excited
states that exist without radiation field turn into resonances. These
are metastable states that decay after a characteristic lifetime.

A reasonable guess for the energies of these resonances to leading order in the coupling $\alpha$ is to compute the expectation value of the Hamiltonian in a state similar to the one we used for the ground state and that has proved to give the right answer to leading order. 
I.e., we consider a state containing only one photon, but replace $\phi_0$ and $e_0$ by some excited state of the unperturbed hydrogen atom with corresponding energy.  
We denote by  $\phi_{n,l}$ an eigenfunction of $\mbox{$\frac 1{2m}$}|p|^2 +V$
with principal quantum number $n$, angular momentum $l$, and
corresponding energy $e_n$ (which is, of course, independent of $l$). Doing this calculation and also the mass renormalization to leading order in $\alpha$, 
it is therefore natural to expect an energy of the metastable excited
states, to leading order in $\alpha$, as
\begin{multline}\label{renbindd}
-\frac{\mph(\beta Z)^2}{2 n ^2} + \frac{\alpha}{\mph} \sum_{j=1}^3 \Big\langle p_j\phi_{n,l}\Big|\,\Ss\left(\frac{\mbox{$\frac 1{2m}$}|p|^2+V-e_{n}}{2\mph}\right) \Big|p_j \phi_{n,l}\Big\rangle  
\\- \frac{\alpha}{\mph} \sum_{i,j=1}^3 \Big\langle p_i\phi_{n,l}\Big|\,\T_{ij}\left(\frac{\mbox{$\frac 1{2m}$}|p|^2+V-e_{n}}{2\mph}, \frac{p}m \right) \Big|p_j \phi_{n,l}\Big\rangle \ .
\end{multline}
Note that the operator $\mbox{$\frac 1{2m}$}|p|^2+V-e_{n}+\mbox{$\frac
1{2m}$}|k|^2+|k|$ is not positive, but nevertheless invertible for
almost every $k$. The expression (\ref{renbindd}) makes perfect sense
if the integrals over $k$ in (\ref{defg}) and (\ref{deftl}) are
interpreted as the Cauchy principal value.

The same discussion as for the ground state energy shift in the
previous section also applies here. Neglecting $\T$ and approximating
$\Ss(e)$ by $\mbox{$\frac 4{3\pi}e\ln(1/e)$}$ gives the result of
Bethe \cite{Be}. This approximation is valid for $Z\ll 1/\alpha$. For
larger $Z$, our formula presumably provides a better description of
the energy level shift due to the presence of the quantized radiation
field than Bethe's formula.

The correctness of the formula for the shift of the energy levels can
best be tested on the classical Lamb shift, namely the splitting of
the $2s_{1/2}$ and $2p_{1/2}$ states, since this is an effect entirely due to the quantized radiation field, and there is no splitting due
to relativistic effects.

\bigskip
\noindent {\it Acknowledgments.} 
We are grateful to Elliott Lieb for helpful discussions.  C.H. was
supported by a Marie Curie Fellowship of the European Community
programme \lq\lq Improving Human Research Potential and the
Socio-economic Knowledge Base\rq\rq\ under contract number
HPMFCT-2000-00660 and by the Deutsche Forschungsgemeinschaft, and acknowledges kind hospitality at Princeton
University, where part of this work was done. R.S. was supported by
the Austrian Science Fund in the form of an Erwin Schr\"odinger
Fellowship.

\end{document}